%% file: integer-splits-paper.tex
\documentclass[a4paper,USenglish,cleveref,autoref,thm-restate,pdfa]{lipics-v2021}

\newcommand{\paraqooba}{\textsc{ParaQooba}}
\newcommand{\slurm}{\textsc{slurm}}
\newcommand{\caqe}{\textsc{CAQE}}
\newcommand{\bloqqer}{\textsc{Bloqqer}}

\newtheorem{assumption}[theorem]{Assumption}

\usepackage{csvsimple}
\usepackage[binary-units]{siunitx}
\usepackage{tikz}
\usepackage[nounderscore]{syntax}
\usepackage{standalone}
\usepackage{pgfplots}
\usepackage{pgfplotstable}
\usepackage{gnuplot-lua-tikz}

\usepackage{microtype}
\makeatletter
\def\MT@is@opt@char#1\iffontchar#2\char#3\else#4\fi\relax{%
  \MT@ifempty{#1}{%
    \iffontchar#2%
      \expandafter\chardef
        \csname\MT@encoding\MT@detokenize@c\@tempa\endcsname=#3\relax
    \fi
  }\relax
}
\makeatother

\def\prefix{\mathcal{Q}}
\def\matrix{\varphi}

\title{Search-Space Pruning with Int-Splits for Faster QBF Solving}

\Copyright{Maximilian Heisinger, Irfansha Shaik, Martina Seidl, and
  Jaco van de Pol} %TODO mandatory, please use full first names. LIPIcs license is "CC-BY";  http://creativecommons.org/licenses/by/3.0/

\authorrunning{M. Heisinger et al.} %TODO mandatory. First: Use abbreviated first/middle names. Second (only in severe cases): Use first author plus 'et al.'


\nolinenumbers %uncomment to disable line numbering


\author{Maximilian Heisinger}%
  {Johannes Kepler University Linz,
    Austria\and\url{https://maxheisinger.at}}%
  {maximilian.heisinger@jku.at}%
  {0000-0001-7297-6000}%
  {}

\author{Irfansha Shaik}%
  {Aarhus University, Denmark\and\url{https://pure.au.dk/portal/en/persons/irfansha-shaik(4dcf450b-e5d7-48ff-8aa0-357484bfe3c7).html}}%
  {irfansha.shaik@cs.au.dk}%
  {0000-0002-7404-348X}%
  {}

\author{Martina Seidl}%
  {Johannes Kepler University Linz,
    Austria\and\url{https://www.jku.at/institut-fuer-symbolic-artificial-intelligence/team/martina-seidl/}}%
  {martina.seidl@jku.at}%
  {0000-0002-3267-4494}%
  {}

\author{Jaco van de Pol}%
  {Aarhus University, Denmark\and\url{https://www.cs.au.dk/~jaco/}}%
  {jaco@cs.au.dk}%
  {0000-0003-4305-0625}%
  {}

\EventEditors{John Q. Open and Joan R. Access}
\EventNoEds{2}
\EventLongTitle{26th International Conference on Theory and Applications of Satisfiability Testing (SAT 2023)}
\EventShortTitle{SAT 2023}
\EventAcronym{SAT}
\EventYear{2023}
\EventDate{July 04-08}
\EventLocation{Alghero, Italy}
\EventLogo{}
\SeriesVolume{42}
\ArticleNo{23}
\begin{document}
\maketitle

\begin{abstract}
  In many QBF encodings, sequences of Boolean variables stand for
  binary representations of integer variables. Examples are state
  labels in bounded model checking or actions in planning problems.
  Often not the full possible range is used, e.g., for representing
  six different states, three Boolean variables are required,
  rendering two of the eight possible assignments irrelevant for the
  solution of the problem. As QBF solvers do not have any
  domain-specific knowledge on the formula they process, they are not
  able to detect this pruning opportunity.

  In this paper, we introduce the idea of \emph{int-splits}, which
  provide domain-specific information on integer variables to QBF
  solvers. This is particularly appealing for parallel
  Divide-and-Conquer solving which partitions the search space into
  independently solvable sub-problems. Using this technique, we reduce
  the number of generated sub-problems from a full expansion to only
  the required subset. We then evaluate how many resources int-splits
  save in problems already well suited for D\&C. In that context, we
  provide a reference implementation that splits QBF formulas into
  many sub-problems with or without int-splits and merges results. We
  finally propose a comment-based optional syntax extension to
  \texttt{(Q)DIMACS} that includes int-splits and is suited for
  supplying proposed guiding paths natively to D\&C solvers.
\end{abstract}

\section{Introduction}

\emph{Quantified Boolean formulas} (QBFs) extend propositional logic
by quantifiers $\forall$, $\exists$ over the Boolean
variables~\cite{DBLP:series/faia/BeyersdorffJLS21}. As a consequence,
the decision problem of QBF (QSAT) is PSPACE complete, making QBF a
suitable framework for many application problems like formal
verification, synthesis, or
planning~\cite{DBLP:conf/ictai/ShuklaBPS19}. In the past years, very
powerful solving tools have been presented that are able to find
solutions even for large instances despite the size of the search
space that is exponential in the number of variables. These tools
mainly exploit structural information within a formula, employ
advanced learning techniques, and implement sophisticated search
strategies (see \cite{DBLP:series/faia/BeyersdorffJLS21} for a recent
survey). Currently, QBF solving tools are completely agnostic of the
problem underlying the encoding, i.e., they do not have any
domain-specific information on the solution space of the original
problem.

For example, the solution to a planning problem is a sequence of
actions to reach a predefined goal~\cite{planning_and_sat}. When
encoding planning problems in propositional logic, a grounding step is
applied, instantiating all action rules with concrete object
combinations. With quantifiers, encodings of planning problems and two-player games become
an order of magnitude more compact compared to the grounded
propositional encoding, decreasing solver memory
requirements~\cite{DBLP:conf/aips/ShaikP22,https://doi.org/10.48550/arxiv.2301.07345}.
% Note: in planning, there is no alternation between quantifiers of subsequent actions 
In game encodings, the first $n$ quantifier alternations encode the
sequence of actions or moves to take in order to reach the specified goal. Each
quantifier block contains a set of variables that, taken together,
represent the action to take in the respective step. Assume a setting
in which $19$ different actions are possible. Then each quantifier
block has to contain five variables. As there are $32$ assignments
possible, $13$ of them are not relevant for the solution. This
information could considerably reduce the search effort. However, so
far it is not available to (or rather not detected by) the solvers.

In this paper, we propose to enrich QBF encodings with domain-specific
information. In particular, we develop a format to restrict the range
of the set of bits that are meant to represent integers. We then use this
information to make divide-and-conquer solvers more efficient and
produce less spurious subproblems, i.e., subproblems that cannot
contribute to the final solution. As case study we consider several
instances from planning problems illustrating the potential of our
approach. 
Our reference implementation is publicly available
under MIT license.

\section{Preliminaries}
\label{sec:prelim}

We consider QBFs $\prefix.\matrix$ in \emph{prenex conjunctive normal
  form} (PCNF) where the \emph{prefix} $\prefix$ is of the form
$Q_1x_1, \ldots, Q_nx_n$ with $Q_i \in \{\forall, \exists\}$. The
\emph{matrix} $\matrix$ is a propositional formula over the variables
$x_1, \ldots, x_n$ in conjunctive normal form (CNF). A formula in CNF
is a conjunction ($\land$) of clauses. A \emph{clause} is a
disjunction ($\lor$) of literals. A literal is a variable $x$, a
negated variable $\lnot x$ or a (possibly negated) truth constant
$\top$ (true) or $\bot$ (false). For a literal $l$, the expression
$\bar l$ denotes $x$ if $l = \neg x$, and it denotes $\lnot x$
otherwise. We sometimes write a clause as a set of literals and a CNF
formula as set of clauses.

If $\matrix$ is a CNF formula, then $\matrix_{x \leftarrow t}$ is the
CNF formula obtained from $\matrix$ by setting $x$ to the value
$t \in \{0, 1\}$. If $t = 0$, then $x$ is set to false, if $t = 1$,
then $x$ is set to true. The formula is then simplified according to
the standard semantics of propositional logic. Given a set $X$ of
propositional variables, then an assignment
$\sigma \colon X \to \{0, 1\}$ maps all variables in $X$ to true or
false. By $\varphi_\sigma$ we denote the formula obtained by setting
all variables which are assigned by $\sigma$.

A QBF $\phi$ is true iff $\nu(\phi) = 1$ where the interpretation 
function $\nu$ that maps QBFs to truth values $\{0, 1\}$ is defined as 
follows: 
\begin{itemize}
	\item $\nu (\forall x\prefix.\matrix) = 1 $ iff 
		$\nu(\prefix.\matrix_{x\leftarrow 0}) \cdot
		\nu(\prefix.\matrix_{x\leftarrow 1}) = 1$
	\item $\nu (\exists x\prefix.\matrix) = 1 $ iff 
		$\nu(\prefix.\matrix_{x\leftarrow 0}) +
		\nu(\prefix.\matrix_{x\leftarrow 1}) \geq 1$
\end{itemize}
Note that we assume all variables of a QBF to be quantified, i.e.,
we are considering closed formulas only. Further, we use standard
semantics of conjunction, disjunction, negation, and truth constants.
For example, the QBF
$\phi_1 = \forall x \exists y. ((x \lor y) \land (\lnot x \lor \lnot
y))$ is true, while
$\phi_2 = \exists y \forall x. ((x \lor y) \land (\lnot x \lor \lnot
y))$ is false. As we see already by this small example, the semantics
impose an ordering on the variables w.r.t.\ the prefix.

\section{Quantified Boolean Formulas with  Int-Splits}

In this section, we extend the language of QBFs with
\emph{int-splits}. With int-splits we provide domain-specific
knowledge to the solver, i.e., we indicate that sets of Boolean
variables stand for integers of a certain range. To this end, we
have to redefine the syntax and semantics of quantifiers, i.e., we
enrich it with information on the integer domain. For the ease of
presentation, we group Boolean variables into bit-vector variables. A
bit-vector variable $\vec{x}$ is a tuple of propositional variables
$(x_1, x_2, \ldots, x_n)$. In this case,
the size $|\vec{x}|$ of $\vec{x}$ is $n$. A bit-vector is a tuple
of bits $\{0, 1\}$. To assign truth values to all bit variables of a
bit-vector of size $n$, we write
${\vec{x} \leftarrow (t_1,t_2,\ldots,t_n)}$ or alternatively
$\sigma(\vec{x})$ if $\sigma$ is an assignment for the propositional
variables occurring in $\vec{x}$. For a CNF formula $\varphi$, the
formula $\varphi_{\sigma(\vec{x})}$ is obtained by setting $x_1$ to
$\sigma(x_1)$, $x_2$ to $\sigma(x_2)$, \ldots, and $x_n$ to
$\sigma(x_n)$. By $\mathcal{S}(\vec{x})$ we denote the set of all
possible assignments to $\vec{x}$, basically representing all
bit-vectors of length $|\vec{x}|$. If $(t_1, \ldots, t_n)$ is a
bit-vector, then its integer representation is
$\langle (t_1, \ldots, t_n) \rangle = 2^{n-1}\cdot{}t_1+ \ldots + 2^0\cdot{}t_n$.
For example if $\vec{x} = (x_1, x_2)$ and $\sigma(x_1) = 1$ and
$\sigma(x_2) = 0$, then
$\langle\sigma(\vec{x})\rangle = \langle (1, 0) \rangle = 2$.

\begin{definition}[Syntax of QBFs with Int-Splits] A QBF with
  int-splits is of the form
  $Q_1(\vec{x_1})_{c_1}\ldots Q_n(\vec{x_n})_{c_n}.\varphi$ where
  $Q_i \in \{\forall, \exists\}$ and $\vec{x_i}$ are bit-vector
  variables of size $\geq 1$ such that there is no Boolean variable
  that occurs in two different bit-vector variables. The constraints
  $c_i \in \{\;{<}v_i,\; {>}v_i,\; {=}V,\; \top\;\}$ restrict the domain of the
  quantifier $Q_i$. In the definition of a constraint
  $v_i \in \mathbb{N^+}$ is a constant and $V$ is a set of
  bit-vectors. As before, $\varphi$ is a propositional formula in CNF
  over the Boolean variables of the $\vec{x_i}$.
\end{definition}

Note that the definition above allows that $Q_i = Q_{i+1}$ and that
the bit-vector variables may be of different sizes. If such a
bit-vector variable is of size one, then it is a simple Boolean
variable. Truth constant $\top$ indicates that there is no restriction
on the range of the preceding bit-vector variable. We often omit the
constraint instead of writing $\top$. Further, note that $\varphi$ is a
propositional formula and not a bit-vector formula. We call
$Q(\vec{x})_c$ an \emph{annotated quantifier}. An example of a QBF
with int-splits is
$$\exists(x_1, x_2)_{<3}\exists(y_1, y_2, y_3)_{=\{101, 111\}}
\forall(z_1, z_2)_{>2}\forall a \exists b.\varphi.$$

\begin{definition}[Semantics of QBFs with Int-Splits] Let
  $\phi = Q\vec{x}_{c}\mathcal{Q}.\varphi$ be a QBF with int-splits
  with constraint $c$ and $Q \in \{\exists, \forall\}$. Then $\phi$ is
  true iff $\nu(\phi) = 1$ where the interpretation function $\nu$ is
  defined as follows.
\begin{itemize}
	\item 
		$\nu (\exists \vec{x}_\top\mathcal{Q}.\varphi) = 1$ iff 
		$\sum_{\sigma\in\mathcal{S}(\vec{x})}\nu(\mathcal{Q}.\varphi_{\sigma(\vec{x})}) \geq 1$
	\item 
		$\nu (\forall \vec{x}_\top\mathcal{Q}.\varphi) = 1$ iff 
		$\prod_{\sigma\in \mathcal{S}(\vec{x})}\nu(\mathcal{Q}.\varphi_{\sigma(\vec{x})}) = 1$
	\item 
		$\nu (\exists\vec{x}_{\diamondsuit v}\mathcal{Q}.\varphi) = 1$ iff 
		$\sum_{\sigma \in \mathcal{S}(\vec{x}), \langle\sigma(\vec{x})\rangle \diamondsuit v}\nu(\mathcal{Q}.\varphi_{\sigma(\vec{x})}) \geq 1$ with $\diamondsuit \in \{<, >\}$

	\item 
		$\nu (\forall \vec{x}_{\diamondsuit v}\mathcal{Q}.\varphi) = 1$ iff 
		$\prod_{\sigma \in \mathcal{S}(\vec{x}), \langle\sigma(\vec{x})\rangle \diamondsuit v}\nu(\mathcal{Q}.\varphi_{\sigma(\vec{x})}) = 1$ with $\diamondsuit \in \{<, >\}$
	
	\item 
		$\nu (\exists\vec{x}_{{=}V}\mathcal{Q}.\varphi) = 1$ iff 
		$\sum_{\sigma \in \mathcal{S}(\vec{x}), \sigma(\vec{x}) \in V}\nu(\mathcal{Q}.\varphi_{\sigma(\vec{x})}) \geq 1$ 
	\item 
		$\nu (\forall\vec{x}_{{=}V}\mathcal{Q}.\varphi) = 1$ iff 
		$\prod_{\sigma \in \mathcal{S}(\vec{x}), \sigma(\vec{x}) \in V}\nu(\mathcal{Q}.\varphi_{\sigma(\vec{x})}) = 1$ 
\end{itemize}

\end{definition}

Note that int-splits correspond to bounded quantifications, which may change the 
semantics of the QBF. This is not our intention: we view int-splits as annotations,
i.e., hints on integer bounds that were already present in the QBF formula, but
which were not recognized by the solver, resulting in an unnecessarily large search
space. It is the responsibility of the user (modeler) to provide correct int-splits, as codified
in the following assumption:

\begin{assumption}[Correctness of Int-Splits] 
\label{assumption}
Let a QBF $\phi$ with int-splits be given and let $\phi'$ 
be the QBF obtained by replacing all int-splits 
constraints from $\phi$ by $\top$.
The int-splits in $\phi$ are \emph{correct} iff $\nu(\phi)=\nu(\phi')$.
From now on, we will assume that the specified int-splits are always correct. 
\end{assumption}

With (correct) int-splits the search space can be considerably reduced, because
we are now able to indicate that there are variable assignments that
cannot contribute to a solution, making certain expansions
unnecessary. Consider the small example in \autoref{fig:int-split}
which shows the formula
$\forall \vec{x}_{<3} \exists \vec{y}_{<3}\prefix.\matrix$ as a tree
of int-splits. Out of 16 branches, nine have to be considered for 
solving, while seven can be ignored. We therefore distinguish between
accounted and unaccounted expansions as follows.

\begin{definition}
  Let $\Phi = \prefix.\matrix$ be a QBF with int-splits containing an
  annotated quantifier $Q_i(\vec{x_i})_{c_i}$. Further let $X$ be the
  set of Boolean variables that occur in $\vec{x_j}$ with $j \leq i$.
  An assignment $\sigma \colon X \to \{0, 1\}$ is an \emph{accounted
    expansion} (AE) for $\vec{x}$ if $c_i(\sigma({\vec{x}}))$ is true.
  Otherwise, $\sigma$ is an \emph{unaccounted expansion} (UE) for
  $\vec{x}$.
\end{definition}

Hence, 
$\vec{x}_{<3}$ and $\exists \vec{y}_{<3}$ have three accounted 
expansions and one unaccounted expansion each. For an 
	annotated quantifier $Q\vec{z}_\top$ the number of 
	AEs is $2^{|\vec{z}|}$, while 
	the number of UEs is zero. In the following, we 
	assume that for each annotated quantifier $Q$
	with $s$ AEs, $s > 0$, because otherwise $Q$ can 
	be removed from the prefix. 

%Further, let $C = \{(\vec{x}, c) \mid 
%Q\vec{x}_c \in \prefix\}$ be the set of constraints in the prefix. 
%	An assignment $\sigma \colon X \to \{0, 1\}$ is an \emph{accounted 
%	expansion} if $c(\sigma({\vec{x}}))$ is true for all 
%	$(\vec{x}, c) \in C$. Otherwise, $\sigma$ is an 
%	\emph{unaccounted expansion} (UE). 

%To give an example, the int-split $[x_1,x_2]_{< 3}$ on the QBF
%$\forall x_1 \forall x_2 \prefix.\matrix$ expands to the AEs
%$\prefix.\matrix_{x_1 \leftarrow \bot,x_2 \leftarrow \bot}$,
%$\prefix.\matrix_{x_1 \leftarrow \bot, x_2 \leftarrow \top}$, and
%$\prefix.\matrix_{x_1 \leftarrow \top ,x_2 \leftarrow \bot}$. The
%assignment
%$\prefix.\matrix_{x_1 \leftarrow \top ,x_2 \leftarrow \top}$ would be
%interpreted as $3$, which does not match the int-split condition as
%$3\geq{}3$ and is therefore a UE. \autoref{fig:int-split} shows this
%split in the context of a D\&C tree. On the other hand,
%\autoref{fig:int-split} shows the same variables with exhaustive
%splitting.

\input{tree}

\section{Int-Splits and Divide and Conquer}

\subsection{Background}
\label{sec:background}

In Divide-and-Conquer (D\&C) SAT
Solving~\cite{DBLP:conf/tacas/FriouxBSK19}, a splitting variable
is selected and two copies of the formula are generated. The
splitting variable is set to true (1) in one copy and to false (0) in the
other copy. This is repeated until a given depth $d$, resulting in $2^d$
subproblems to be solved. While in SAT the decision variable can be
freely selected, the possible choices for variable selection are more
restricted for a QBF $Q_1XQ_2Y\prefix.\matrix$ with $Q_1 \not= Q_2$
and $Q_1, Q_2 \in \{\forall, \exists\}$. Only variables from the
outermost quantifier block $Q_1 X$ may be considered for the splitting
variable and only if all variables from $X$ are assigned, variables from
$Y$ may be considered. While in SAT solving, the solving process could
be stopped as soon as one sub-problem is found to be satisfiable, in
QBF solving, the results of the subproblems need to be merged taking
the different quantifier types into account.

In SAT, the D\&C approach has been especially successful for hard
combinatorial SAT problems in a variant called Cube-and-Conquer
(C\&C)~\cite{DBLP:conf/hvc/HeuleKWB11}. While the latter relies on
advanced heuristics to select the splitting variables, Jordan et
al.~\cite{DBLP:conf/sat/JordanKLS14} observed that the sequential
order of variables in the first quantifier block already leads to
improvements in the case of QBF solving. Heisinger et
al.~\cite{DBLP:conf/tacas/Heisinger23} combined Portfolio solving with
D\&C Solving for QBF and showed further improvements. Here a 
rudimentary version of int-splits was suggested in terms of 
a command line option \texttt{-{}-intsplit} $n$ that creates 
$n$ tasks over the first $\left \lceil{\log_2 n}\right \rceil$
variables. In contrast to the approach presented in this paper, 
there is no connection to the encoding, the expressible constraints 
are more restricted, and there is no optimization w.r.t.\ efficiency
$\eta$. 
Unpublished experiments with
\paraqooba{} indicated the potential for int-splits (but also 
the danger of incomplete solving), which led to this work
to formalize, extend, and evaluate int-splits in-depth.

\subsection{Combining D\&C with Int-Splits}
\label{sec:gener-dc-using}

While for $d$ splitting variables $2^d$ sub-problems have to be
generated by a D\&C solver, int-splits have the potential to
substantially decrease the number of sub-problems to be evaluated. Let
$Q(\vec{x_1})_{c_1}\ldots Q(\vec{x_n})_{c_n}\prefix.\matrix$ be a QBF
with int-splits and $|\vec{x_1}| + \ldots + |\vec{x_n}| = d$.\footnote{
	$|\vec{x_1}| + \ldots + |\vec{x_{n-1}}| < d <
	|\vec{x_1}| + \ldots + |\vec{x_{n}}|$, int-splits are only applied on 
	the first $n-1$ bit-vectors. }
	Further,
let $s_i = |\{ c_i(\sigma(\vec{x_i})) \mid \sigma \in S(\vec{x_i}\}|$,
i.e., the number of AEs of $\vec{x_i}$. Then the number of unaccounted
expansions of $\vec{x_i}$ is $u_i = 2^{|\vec{x}|} - s_i$. Instead of
$2^d$ sub-problems, with int-splits $s_1 \cdot \ldots \cdot s_n$ problems need
to be considered.

\begin{example}
  Consider the formula
  $Q_1(\vec{x_1})_{<19}Q_2(\vec{x_2})_{<19}Q_2(\vec{x_2})_{<19}\prefix.\matrix$
  with $|\vec{x_i}| = 5$ and a splitting depth of 15. In conventional
  D\&C solving, $2^5 \cdot 2^5 \cdot 2^5= 2^{15} = 32768$ sub-problems
  would be generated. In contrast, with the int-split information,
  each $x_i$ has 19 AEs (and therefore 13 UEs). Overall
  $19 \cdot 19 \cdot 19 = 6859$ have to be generated, which is $\approx 21\%$
  of the full expansion.
\end{example}

When a prefix has the form $Q(\vec{x_1})_{c_1}\ldots Q(\vec{x_n})_{c_n}\prefix.\matrix$ with $Q \in \{\forall, \exists\}$, but the splitting depth 
$d < |\vec{x_1}| + \ldots + |\vec{x_n}|$, then the expansions with the 
largest pruning potential should be selected for expansion. To this end, we 
introduce the notion of int-split efficiency that allows to 
reorder annotated quantifiers of the same kind.

\iffalse
\begin{figure}[t]
  \centering
  \includestandalone[mode=buildmissing]{efficiency}
  \caption{$\eta$ for varying $m$ and different int-split layer
    counts.}
  \label{fig:Meff}
\end{figure}
\fi

\begin{definition}[Int-Split Efficiency]
  Let $\Phi = \prefix.\matrix$ be a QBF with int-splits containing an
  annotated quantifier $Q(\vec{x})_{c}$ which 
  has $s$ AEs and $u$ UEs. Then 
  $Q(\vec{x})_{c}$ has \emph{int-split efficiency} $\eta$ 
	with $\eta = \frac{u}{s}$.  
  \label{def:int-split-efficiency}
\end{definition}

Recall that for each annotated quantifier $Q(\vec{x})_c$ 
with $s$ AEs, it holds that $s > 0$. For example 
$Q((x_1,x_2))_{<3}$ has $\eta = \frac{1}{3}$ and 
$Q(\vec{y})_\top$ has $\eta = 0$. 

\section{Reference Implementation}

In this section, we introduce our reference implementation for testing
the viability of int-splits in a D\&C context. First, we propose a
comment-based back- and forwards compatible format extension to the
(Q)DIMACS file format.\footnote{http://www.qbflib.org/qdimacs.html}  
Afterwards, we describe how we split problems
into sub-problems under the constraints of the 
annotated quantifiers. Finally,  we explain
how we merge results of sub-problems to obtain a solution of 
the original formula. 
Our reference implementation is MIT-licensed and available publicly at
\begin{center}
\href{https://github.com/maximaximal/qdimacs-splitter}{github.com/maximaximal/qdimacs-splitter}
\end{center}
It is written in Rust and highly scalable w.r.t.\ formula size. 
While reading int-splits of the same quantification 
type, it sorts them with decreasing
efficiency $\eta$ locally to each annotated quantifier. 
It expands formulas along the sorted prefix
into copies of the original formulas, with assumptions appended as unit
clauses. If an expanded
variable is universally quantified, its quantifier is changed to 
existential in the copy 
for the application of Boolean constraint propagation. 
Every copy is saved as a separate file, named after
the original formula with the index of the respective copy prepended.

\subsection{(Q)DIMACS Format Extension}

We annotate formulas in (Q)DIMACS with int-splits formulated in 
as comments in order to preserve compatibility with 
the (Q)DIMACS format. 
We include the extended 
grammar in~\autoref{sec:qdimacs-extension-grammar}. 
While the 
int-split annotations and the prefix have to match in terms of 
quantifier blocks, orderings within quantifier blocks might 
be different. 
Note that without the correctness assumption formulated 
above, the semantics of the QBF might change. Currently, it is 
in the responsibility of the user to employ the annotations correctly. 
In our implementation, an annotated quantifier may include multiple 
constraints separated by semicolons. Then an expansion is accounted, 
if least one of the constraints is satisfied. 
If the number $s$ of AEs can be uniquely calculated, then
$\vec{x}$ can be left empty, indicating that the next 
$\lceil{}\log_2 s\rceil{}$ variables from the prefix
have to be considered. To generate QDIMACS a prefix must be given, to
generate DIMACS, all $\vec{x}$ must be explicit and no prefix is allowed.

\subsection{Splitting Formulas}

For computing the required splits, our tool tracks the number of
Boolean variables to split over, creates a generator over all possible
assignments to the variables, iterates over it and checks for every
possible assignment, if all considered annotated quantifiers
have at least one of their
constraints satisfied. The splitting depth parameter \texttt{-{}-depth}
defines the upper bound on the number of variables to be considered
for splitting, taking as many (int-) splits as possible. For any given
splitting depth $d$, at most $2^d$ files are created, each with the
same content as the original formula, save for the appended
unit-clauses for the assumptions. As the splitting and merging are
decoupled from the (potentially parallel) solving itself, we introduce
this D\&C variant as \emph{offline splitting}.

\subsection{Merging Results}
\label{sec:merging}

\emph{Offline splitting} enables the use of parallel execution tools
like e.g. Simsala~\cite{simsala}, while also allowing all expansions
to be run sequentially, only simulating a parallel execution. After a
solver run has finished, our merger reads the original formula and
$n$ annotated quantifiers that have been expanded in memory. Then,
it reads all logs and maps the files to $r_n = \prod^n s_i$ tuples of result
codes and runtimes $(r^n_j,t^n_j)$ with $1 \leq j \leq r_n$. 
These $r_n$ tuples are reduced to $r_{n-1} = \prod^{n-1} s_i$ tuples
by taking the maximal result value and the minimal time value if $Q_n = \exists$ and the minimum result value and the maximal time value if $Q_n = \forall$. 
This process
repeats, until the new list only contains one result, representing the
root of the tree with the final result and the wall-clock time of 
a virtual parallel solver with $r_n$ processors or a sequential solver 
with optimal decision heuristics. 
%The merger computes the shortest path to the final solution that also
%resembles the wall-clock time of parallel solvers. If a D\&C parallel
%solver either has enough cores available to solve all sub-problems at
%once, or if its decision heuristic is optimal and makes no mistakes,
%it's wall-clock runtime will at best be equal to the shortest path
%computed by our merging tool. 
Our merger is thus suitable for
researching decision heuristics, as the result table it generates
provides an efficiency upper-bound for improving heuristic functions,
without being limited by the physical number of available cores or
difficulties with parallel programming itself. The internal
representation of sub-problems mapped to results and runtimes can be
interpreted as a certificate of parallel execution.

\section{Evaluation}

We expanded the planning benchmarks used
in~\cite{DBLP:conf/tacas/Heisinger23} with int-splits and compared the
results to the same formulas without int-splits. All sub-formulas were
run with a timeout of \SI{3700}{\second} on a cluster of dual-socket
AMD EPYC 7313 @ 16x \SI{3.7}{\giga\hertz} machines with
\SI{8}{\giga\byte} memory per task. We benchmarked using the QBF
solver \caqe{}~\cite{DBLP:conf/fmcad/RabeT15} and pre-process the
split formulas with \bloqqer{}~\cite{DBLP:conf/cade/BiereLS11}, as
this benefits the selected formulas the most. Both the Hex and GTTT
formulas have a prefix of the form
$\exists(x_1x_2x_3x_4x_5)_{<19}\forall(x_6x_7x_8x_9x_{10})_{<19}\exists(x_{11}x_{12}x_{13}x_{14}x_{15})_{<19}\ldots$.

\subsection{CPU-Time Comparison}
\label{sec:cpu-time-comparison}

\begin{figure}[b]
  \centering
  \begin{minipage}{0.31\linewidth}
  \includestandalone[mode=buildmissing,width=\textwidth]{scatter}%
  \end{minipage}%
  \begin{minipage}{0.69\linewidth}
  \includestandalone[mode=buildmissing,width=\textwidth]{sn-gttt}
  \includestandalone[mode=buildmissing,width=\textwidth]{sn-hex}
  \end{minipage}
  \caption{CPU Time taken to process the same formula with and without
    int-splits. Data includes Satisfiable, Unsatisfiable, and Timeout
    (all with $t>500$) results. The y-axis is fit to the runtime of
    the solved instances. Top barchart contains GTTT results, bottom
    Hex.}
  \label{fig:cpu-time-scatter}
\end{figure}

% Table with interesting data of formulas.

\begin{table}[t]
  \centering
  \resizebox{\textwidth}{!}{
  \begin{tabular}{l%
    S[table-format=1]%
    S[table-format=1]%
    S[table-format=1]%
    S[table-format=1]%
    S[table-format=1]%
    S[table-format=1]%
    S[table-format=1,round-mode=off]%
    S[table-format=3,round-mode=off]%
    S[table-format=3,round-mode=figures]%
    S[table-format=3.2,table-column-width=1.2cm,round-mode=figures,round-precision=3]%
    S[table-format=3.2,table-column-width=1.2cm,round-mode=figures,round-precision=3]%
    S[table-format=3.2,round-mode=figures]%
    S[table-format=3.2,table-column-width=1.2cm,round-mode=figures,round-precision=3]}
    {Formula}%
    & {\#v}%
    & {\#c}%
    & {$s$}%
    & {$u$}%
    & {$d_{\text{w/}}$}%
    & {$d_{\text{w/o}}$}%
    & {$\#_{\text{w/}}$}%
    & {$\#_{\text{w/o}}$}%
    & {$\eta$}%
    & {$t_{\text{w/}}$ [h]}%
    & {$t_{\text{w/o}}$ [h]}%
    & {$\frac{t_{\text{w/}}}{t_{\text{w/o}}}$}
    & {$\frac{t_{\text{seq}}}{t_{\text{par}}}$} \\
    % & Result \\
    \hline
    \csvreader[respect sharp=true,respect underscore=true,no head,%
    before filter=\ifthenelse{\equal{\csvcoli}{Formula}}{\csvfilterreject}{\csvfilteraccept}]%
    {overview-table.csv}{}{%
    \csvcoli & \csvcolii & \csvcoliii & \csvcoliv & \csvcolv%
    & \csvcolvi & \csvcolvii & \csvcolviii & \csvcolix & \csvcolx
    & \csvcolxi & \csvcolxii & \csvcolxiii & \csvcolxix \\}
  \end{tabular}
  }
  \caption{Selected benchmark results without timeouts highlighting
    differences between solving with (w/) and without (w/o)
    int-splits. All formulas have quantifier alternations of the form
    $\exists\forall\ldots$. The GTTTs and Hex 15 5x5-15 have 15
    alternations and the other Hex formulas 13. $d_{\text{w/}}$ is
    lower than $d_{\text{w/o}}$ if no more int-splits could be fit
    into the given depth $d=10$. $t$ is total CPU time. \# is the
    total number of generated sub-problems. The less $u$, the closer
    $\frac{t_{\text{w/}}}{t_{\text{w/o}}}$ gets to $1$. The right-most
    column $\frac{t_{\text{seq}}}{t_{\text{par}}}$ is the speed-up of
    optimal parallel solving (see~\autoref{sec:merging})
    compared to solving the formula sequentially. They are equal with
    and without int-splits, but int-splits drastically reduce the
    required resources (CPU-time).}
  \label{tab:results}
\end{table}

Our experiment compares how much CPU-time is needed to solve a
number of formulas, once with int-splits and once without. We added
int-split definitions to the formulas and split with $d=10$. This
produces at most $2^{10}=1024$ sub-problems. Depending on the
int-split definition in the formula, exponentially less sub-problems
are produced when splitting with enabled int-splits. We tested
on lifted QBF encodings of planning problems in Organic Synthesis (obtained from ~\cite{DBLP:conf/aips/ShaikP22})
and on lifted QBF encodings of the 2-player positional games Hex and generalized Tic-Tac-Toe (obtained from~\cite{https://doi.org/10.48550/arxiv.2301.07345}). 
The results in~\autoref{fig:cpu-time-scatter} follow the expectation of
less CPU time with int-splits.

In order to have a fair comparison, we only include results where the
maximum depth $d$ is equal to the number of variables covered by
int-splits when splitting with $d$. This means that the number of AEs
with int-splits is equal to the number of AEs without int-splits, and
only UEs increase the CPU-time further. Otherwise, int-splits results
would cover less AEs and the effective splitting depth with int-splits
would be lower than without. This is illustrated by the last row
in~\autoref{tab:results}, which was left in to show this effect.

\subsection{Wall-Clock Time Comparison}
\label{sec:wall-time-comparison}

We also computed the wall-clock time for the benchmarks above.  
The observed speed-ups with int-splits compared to
sequentially solving the whole formulas are higher than the speed-ups
observed with \paraqooba{} without int-splits, as we could
double the splitting depth. Speed-ups with offline splitting are as
high as 47.7, while the highest speed-up of the online-splitting
solver \paraqooba{} (on a single 16 core node) is 18.09. In contrast
to \paraqooba{}, our splitter and merger runs individual tasks on a
cluster and can therefore be scheduled more efficiently than big
parallel benchmarks. Offline splitting does not currently abort the
solving process when a solution is found, although such integration
into the runtime environment would be scriptable (using e.g.
\slurm{}'s \texttt{scancel} command).

\section{Conclusion}

We presented a language-extension to QBFs in PCNF to support solvers with 
domain-specific knowledge in terms of bounded quantifiers. 
While we have clearly demonstrated the potential of int-splits,
this work opens several interesting questions for future work.
Currently, it is the responsibility of the user to provide correct
int-splits. One question is how to validate efficiently that the
int-splits are correct (by full verification or by a testing approach).
Another question is under which circumstances int-splits can
be inferred automatically. In both cases, Assumption~\ref{assumption}
would be guaranteed by the tools.

Other perspectives for future work
are to improve QBF pre-processors and back-end QBF solvers
to exploit the int-split information as well. Also, int-splits can be easily
used to implement certain forms of symmetry reduction, by
restricting the bounds for some of the variables.
Finally, if all tools in the chain
are aware of the int-splits, Assumption~\ref{assumption} would 
not be necessary anymore. Instead, in that case, one might view 
the int-splits as integral
part of the QBF encoding, and simplify the matrix of the QBF formula 
(i.e., remove the bound-checks from the encoding).

\section*{Data Availability}

Data used for benchmarking the described software, including source
code, are made available permanently under a permissive license in a
public artifact on Zenodo. Raw source data for the figures presented
in this paper are also included~\cite{heisinger_maximilian_2023_zenodo}.

\bibliography{integer-splits-paper}

\newpage
\appendix

\section{Grammar of our (Q)DIMACS Syntax Extension}
\label{sec:qdimacs-extension-grammar}

\grammarindent1.5in
\paragraph*{Preamble and Int-Splits Entry-Point}
\begin{grammar}
  <input> ::= <preamble> [<prefix>] <matrix> \texttt{EOF}
  \vspace{1.2em}

  <preamble>      ::= <comment or split>* <problem line>
  
  <comment or split>  ::= `c' <text> \texttt{EOL}
  \alt `cs' <split> \texttt{EOL}
  
  <problem line>  ::= `p cnf' <pnum> <pnum> \texttt{EOL}

\end{grammar}
\paragraph*{Our Int-Splits Syntax Extension}
\begin{grammar}

  <split>  ::= int [ <int-split-vars> ] <int-split-list>

  <int-split-vars> ::= `[' <pnum>+ `]'
  
  <int-split-list> ::= <int-split> [ `;' <int-split-list> ]

  <int-split> ::= `\(<\)' <pnum>
  \alt `\(>\)' <pnum>
  \alt `\(=\)' `{' <bit pattern>+ `}'
\end{grammar}
\paragraph*{Optional Quantifier Prefix}
\begin{grammar}

  <prefix>     ::= <quant set>+
  
  <quant set>  ::= <quantifier> <pnum>+ `0' \texttt{EOL}
  
  <quantifier> ::= `e' | `a'
\end{grammar}
\paragraph*{Clause Matrix}
\begin{grammar}

  <matrix>      ::= <clause>*
  
  <clause>      ::= <literal>* `0' \texttt{EOL}
  
  <literal>     ::= <num>
\end{grammar}
\paragraph*{Definitions}
\begin{grammar}

  <text>  ::= {\textit{A sequence of non-special ASCII characters}}
  
  <num>   ::= {\textit{A 32-bit signed integer different from 0}}
  
  <pnum>  ::= {\textit{A 32-bit signed integer greater than 0}}

  <bit pattern>  ::= {\textit{A max. 32-bit long pattern of 0 and 1
      without spaces}}
\end{grammar}

\end{document}

%% file: tree.tex
\begin{figure}[t]
  \centering
  \begin{tikzpicture}[grow'=right,level distance=5cm,
    level 1/.style={sibling distance=3cm},
    level 2/.style={sibling distance=0.75cm}]
    \node[draw](z){$\forall{}\vec{x}_{<3}\exists{}\vec{y}_{<3}\prefix.\matrix$}
    child{node[draw]{$\exists{}\vec{y}_{<3}\prefix.\matrix_{\vec{x} \leftarrow (0,0)}$}
      child{node[draw] {
          $\prefix.\matrix_{\vec{x} \leftarrow (0,0),\vec{y} \leftarrow (0,0)}$}
      }
      child{node[draw] {
          $\prefix.\matrix_{\vec{x} \leftarrow (0,0),\vec{y} \leftarrow (0,1)}$}
      }
      child{node[draw] {
          $\prefix.\matrix_{\vec{x} \leftarrow (0,0),\vec{y} \leftarrow (1,0)}$}
      }
      child{node[draw,dashed] {
          $\prefix.\matrix_{\vec{x} \leftarrow (0,0),\vec{y} \leftarrow (1,1)}$}
      }
    }
    child{node[draw]{$\exists{}\vec{y}_{<3}\prefix.\matrix_{\vec{x} \leftarrow (0,1)}$}
      child{node[draw] {
          $\prefix.\matrix_{\vec{x} \leftarrow (0,1),\vec{y} \leftarrow (0,0)}$}
      }
      child{node[draw] {
          $\prefix.\matrix_{\vec{x} \leftarrow (0,1),\vec{y} \leftarrow (0,1)}$}
      }
      child{node[draw] {
          $\prefix.\matrix_{\vec{x} \leftarrow (0,1),\vec{y} \leftarrow (1,0)}$}
      }
      child{node[draw,dashed] {
          $\prefix.\matrix_{\vec{x} \leftarrow (0,1),\vec{y} \leftarrow (1,1)}$}
     }
    }
    child{node[draw]{$\exists{}\vec{y}_{<3}\prefix.\matrix_{\vec{x} \leftarrow (1,0)}$}
      child{node[draw] {
          $\prefix.\matrix_{\vec{x} \leftarrow (1,0),\vec{y} \leftarrow (0,0)}$}
      }
      child{node[draw] {
          $\prefix.\matrix_{\vec{x} \leftarrow (1,0),\vec{y} \leftarrow (0,1)}$}
      }
      child{node[draw] {
          $\prefix.\matrix_{\vec{x} \leftarrow (1,0),\vec{y} \leftarrow (1,0)}$}
      }
      child{node[draw,dashed] {
          $\prefix.\matrix_{\vec{x} \leftarrow (1,0),\vec{y} \leftarrow (1,1)}$}
      }
    }
    child{node[draw,dashed]{$\exists{}\vec{y}_{<3}\prefix.\matrix_{\vec{x} \leftarrow (1,1)}$}
      child{node[draw,dashed] {
          $\prefix.\matrix_{\vec{x} \leftarrow (1,1),\vec{y} \leftarrow (0,0)}$}
      }
      child{node[draw,dashed] {
          $\prefix.\matrix_{\vec{x} \leftarrow (1,1),\vec{y} \leftarrow (0,1)}$}
      }
      child{node[draw,dashed] {
          $\prefix.\matrix_{\vec{x} \leftarrow (1,1),\vec{y} \leftarrow (1,0)}$}
      }
      child{node[draw,dashed] {
          $\prefix.\matrix_{\vec{x} \leftarrow (1,1),\vec{y} \leftarrow (1,1)}$}
      }
    };
  \end{tikzpicture}
  \caption{Two layers of int-splits. The dashed boxes are non-expanded
    UEs ($7/16\approx 44\%$).}
  \label{fig:int-split}
\end{figure}

%% file: scatter.tex
\begin{tikzpicture}[gnuplot]
%% generated with GNUPLOT 5.4p1 (Lua 5.3; terminal rev. Jun 2020, script rev. 114)
%% Mon Mar 20 12:44:48 2023
\gpcolor{color=gp lt color axes}
\gpsetlinetype{gp lt axes}
\gpsetdashtype{gp dt axes}
\gpsetlinewidth{0.50}
\draw[gp path] (1.504,1.366)--(6.447,1.366);
\gpcolor{color=gp lt color border}
\gpsetlinetype{gp lt border}
\gpsetdashtype{gp dt solid}
\gpsetlinewidth{1.00}
\draw[gp path] (1.504,1.366)--(1.684,1.366);
\draw[gp path] (6.447,1.366)--(6.267,1.366);
\node[gp node right] at (1.320,1.366) {$0.1$};
\draw[gp path] (1.504,1.738)--(1.594,1.738);
\draw[gp path] (6.447,1.738)--(6.357,1.738);
\draw[gp path] (1.504,1.956)--(1.594,1.956);
\draw[gp path] (6.447,1.956)--(6.357,1.956);
\draw[gp path] (1.504,2.110)--(1.594,2.110);
\draw[gp path] (6.447,2.110)--(6.357,2.110);
\draw[gp path] (1.504,2.230)--(1.594,2.230);
\draw[gp path] (6.447,2.230)--(6.357,2.230);
\draw[gp path] (1.504,2.328)--(1.594,2.328);
\draw[gp path] (6.447,2.328)--(6.357,2.328);
\draw[gp path] (1.504,2.411)--(1.594,2.411);
\draw[gp path] (6.447,2.411)--(6.357,2.411);
\draw[gp path] (1.504,2.482)--(1.594,2.482);
\draw[gp path] (6.447,2.482)--(6.357,2.482);
\draw[gp path] (1.504,2.545)--(1.594,2.545);
\draw[gp path] (6.447,2.545)--(6.357,2.545);
\gpcolor{color=gp lt color axes}
\gpsetlinetype{gp lt axes}
\gpsetdashtype{gp dt axes}
\gpsetlinewidth{0.50}
\draw[gp path] (1.504,2.602)--(6.447,2.602);
\gpcolor{color=gp lt color border}
\gpsetlinetype{gp lt border}
\gpsetdashtype{gp dt solid}
\gpsetlinewidth{1.00}
\draw[gp path] (1.504,2.602)--(1.684,2.602);
\draw[gp path] (6.447,2.602)--(6.267,2.602);
\node[gp node right] at (1.320,2.602) {$1$};
\draw[gp path] (1.504,2.974)--(1.594,2.974);
\draw[gp path] (6.447,2.974)--(6.357,2.974);
\draw[gp path] (1.504,3.192)--(1.594,3.192);
\draw[gp path] (6.447,3.192)--(6.357,3.192);
\draw[gp path] (1.504,3.346)--(1.594,3.346);
\draw[gp path] (6.447,3.346)--(6.357,3.346);
\draw[gp path] (1.504,3.466)--(1.594,3.466);
\draw[gp path] (6.447,3.466)--(6.357,3.466);
\draw[gp path] (1.504,3.564)--(1.594,3.564);
\draw[gp path] (6.447,3.564)--(6.357,3.564);
\draw[gp path] (1.504,3.647)--(1.594,3.647);
\draw[gp path] (6.447,3.647)--(6.357,3.647);
\draw[gp path] (1.504,3.718)--(1.594,3.718);
\draw[gp path] (6.447,3.718)--(6.357,3.718);
\draw[gp path] (1.504,3.781)--(1.594,3.781);
\draw[gp path] (6.447,3.781)--(6.357,3.781);
\gpcolor{color=gp lt color axes}
\gpsetlinetype{gp lt axes}
\gpsetdashtype{gp dt axes}
\gpsetlinewidth{0.50}
\draw[gp path] (1.504,3.838)--(6.447,3.838);
\gpcolor{color=gp lt color border}
\gpsetlinetype{gp lt border}
\gpsetdashtype{gp dt solid}
\gpsetlinewidth{1.00}
\draw[gp path] (1.504,3.838)--(1.684,3.838);
\draw[gp path] (6.447,3.838)--(6.267,3.838);
\node[gp node right] at (1.320,3.838) {$10$};
\draw[gp path] (1.504,4.210)--(1.594,4.210);
\draw[gp path] (6.447,4.210)--(6.357,4.210);
\draw[gp path] (1.504,4.428)--(1.594,4.428);
\draw[gp path] (6.447,4.428)--(6.357,4.428);
\draw[gp path] (1.504,4.582)--(1.594,4.582);
\draw[gp path] (6.447,4.582)--(6.357,4.582);
\draw[gp path] (1.504,4.702)--(1.594,4.702);
\draw[gp path] (6.447,4.702)--(6.357,4.702);
\draw[gp path] (1.504,4.800)--(1.594,4.800);
\draw[gp path] (6.447,4.800)--(6.357,4.800);
\draw[gp path] (1.504,4.883)--(1.594,4.883);
\draw[gp path] (6.447,4.883)--(6.357,4.883);
\draw[gp path] (1.504,4.954)--(1.594,4.954);
\draw[gp path] (6.447,4.954)--(6.357,4.954);
\draw[gp path] (1.504,5.017)--(1.594,5.017);
\draw[gp path] (6.447,5.017)--(6.357,5.017);
\gpcolor{color=gp lt color axes}
\gpsetlinetype{gp lt axes}
\gpsetdashtype{gp dt axes}
\gpsetlinewidth{0.50}
\draw[gp path] (1.504,5.074)--(6.447,5.074);
\gpcolor{color=gp lt color border}
\gpsetlinetype{gp lt border}
\gpsetdashtype{gp dt solid}
\gpsetlinewidth{1.00}
\draw[gp path] (1.504,5.074)--(1.684,5.074);
\draw[gp path] (6.447,5.074)--(6.267,5.074);
\node[gp node right] at (1.320,5.074) {$100$};
\draw[gp path] (1.504,5.446)--(1.594,5.446);
\draw[gp path] (6.447,5.446)--(6.357,5.446);
\draw[gp path] (1.504,5.664)--(1.594,5.664);
\draw[gp path] (6.447,5.664)--(6.357,5.664);
\draw[gp path] (1.504,5.818)--(1.594,5.818);
\draw[gp path] (6.447,5.818)--(6.357,5.818);
\draw[gp path] (1.504,5.938)--(1.594,5.938);
\draw[gp path] (6.447,5.938)--(6.357,5.938);
\draw[gp path] (1.504,6.036)--(1.594,6.036);
\draw[gp path] (6.447,6.036)--(6.357,6.036);
\draw[gp path] (1.504,6.119)--(1.594,6.119);
\draw[gp path] (6.447,6.119)--(6.357,6.119);
\draw[gp path] (1.504,6.190)--(1.594,6.190);
\draw[gp path] (6.447,6.190)--(6.357,6.190);
\draw[gp path] (1.504,6.253)--(1.594,6.253);
\draw[gp path] (6.447,6.253)--(6.357,6.253);
\gpcolor{color=gp lt color axes}
\gpsetlinetype{gp lt axes}
\gpsetdashtype{gp dt axes}
\gpsetlinewidth{0.50}
\draw[gp path] (1.504,6.310)--(6.447,6.310);
\gpcolor{color=gp lt color border}
\gpsetlinetype{gp lt border}
\gpsetdashtype{gp dt solid}
\gpsetlinewidth{1.00}
\draw[gp path] (1.504,6.310)--(1.684,6.310);
\draw[gp path] (6.447,6.310)--(6.267,6.310);
\node[gp node right] at (1.320,6.310) {$1000$};
\gpcolor{color=gp lt color axes}
\gpsetlinetype{gp lt axes}
\gpsetdashtype{gp dt axes}
\gpsetlinewidth{0.50}
\draw[gp path] (1.504,1.366)--(1.504,6.310);
\gpcolor{color=gp lt color border}
\gpsetlinetype{gp lt border}
\gpsetdashtype{gp dt solid}
\gpsetlinewidth{1.00}
\draw[gp path] (1.504,1.366)--(1.504,1.546);
\draw[gp path] (1.504,6.310)--(1.504,6.130);
\node[gp node center] at (1.504,1.058) {$0.1$};
\draw[gp path] (1.876,1.366)--(1.876,1.456);
\draw[gp path] (1.876,6.310)--(1.876,6.220);
\draw[gp path] (2.094,1.366)--(2.094,1.456);
\draw[gp path] (2.094,6.310)--(2.094,6.220);
\draw[gp path] (2.248,1.366)--(2.248,1.456);
\draw[gp path] (2.248,6.310)--(2.248,6.220);
\draw[gp path] (2.368,1.366)--(2.368,1.456);
\draw[gp path] (2.368,6.310)--(2.368,6.220);
\draw[gp path] (2.466,1.366)--(2.466,1.456);
\draw[gp path] (2.466,6.310)--(2.466,6.220);
\draw[gp path] (2.548,1.366)--(2.548,1.456);
\draw[gp path] (2.548,6.310)--(2.548,6.220);
\draw[gp path] (2.620,1.366)--(2.620,1.456);
\draw[gp path] (2.620,6.310)--(2.620,6.220);
\draw[gp path] (2.683,1.366)--(2.683,1.456);
\draw[gp path] (2.683,6.310)--(2.683,6.220);
\gpcolor{color=gp lt color axes}
\gpsetlinetype{gp lt axes}
\gpsetdashtype{gp dt axes}
\gpsetlinewidth{0.50}
\draw[gp path] (2.740,1.366)--(2.740,6.310);
\gpcolor{color=gp lt color border}
\gpsetlinetype{gp lt border}
\gpsetdashtype{gp dt solid}
\gpsetlinewidth{1.00}
\draw[gp path] (2.740,1.366)--(2.740,1.546);
\draw[gp path] (2.740,6.310)--(2.740,6.130);
\node[gp node center] at (2.740,1.058) {$1$};
\draw[gp path] (3.112,1.366)--(3.112,1.456);
\draw[gp path] (3.112,6.310)--(3.112,6.220);
\draw[gp path] (3.329,1.366)--(3.329,1.456);
\draw[gp path] (3.329,6.310)--(3.329,6.220);
\draw[gp path] (3.484,1.366)--(3.484,1.456);
\draw[gp path] (3.484,6.310)--(3.484,6.220);
\draw[gp path] (3.604,1.366)--(3.604,1.456);
\draw[gp path] (3.604,6.310)--(3.604,6.220);
\draw[gp path] (3.701,1.366)--(3.701,1.456);
\draw[gp path] (3.701,6.310)--(3.701,6.220);
\draw[gp path] (3.784,1.366)--(3.784,1.456);
\draw[gp path] (3.784,6.310)--(3.784,6.220);
\draw[gp path] (3.856,1.366)--(3.856,1.456);
\draw[gp path] (3.856,6.310)--(3.856,6.220);
\draw[gp path] (3.919,1.366)--(3.919,1.456);
\draw[gp path] (3.919,6.310)--(3.919,6.220);
\gpcolor{color=gp lt color axes}
\gpsetlinetype{gp lt axes}
\gpsetdashtype{gp dt axes}
\gpsetlinewidth{0.50}
\draw[gp path] (3.976,1.366)--(3.976,6.310);
\gpcolor{color=gp lt color border}
\gpsetlinetype{gp lt border}
\gpsetdashtype{gp dt solid}
\gpsetlinewidth{1.00}
\draw[gp path] (3.976,1.366)--(3.976,1.546);
\draw[gp path] (3.976,6.310)--(3.976,6.130);
\node[gp node center] at (3.976,1.058) {$10$};
\draw[gp path] (4.347,1.366)--(4.347,1.456);
\draw[gp path] (4.347,6.310)--(4.347,6.220);
\draw[gp path] (4.565,1.366)--(4.565,1.456);
\draw[gp path] (4.565,6.310)--(4.565,6.220);
\draw[gp path] (4.719,1.366)--(4.719,1.456);
\draw[gp path] (4.719,6.310)--(4.719,6.220);
\draw[gp path] (4.839,1.366)--(4.839,1.456);
\draw[gp path] (4.839,6.310)--(4.839,6.220);
\draw[gp path] (4.937,1.366)--(4.937,1.456);
\draw[gp path] (4.937,6.310)--(4.937,6.220);
\draw[gp path] (5.020,1.366)--(5.020,1.456);
\draw[gp path] (5.020,6.310)--(5.020,6.220);
\draw[gp path] (5.091,1.366)--(5.091,1.456);
\draw[gp path] (5.091,6.310)--(5.091,6.220);
\draw[gp path] (5.155,1.366)--(5.155,1.456);
\draw[gp path] (5.155,6.310)--(5.155,6.220);
\gpcolor{color=gp lt color axes}
\gpsetlinetype{gp lt axes}
\gpsetdashtype{gp dt axes}
\gpsetlinewidth{0.50}
\draw[gp path] (5.211,1.366)--(5.211,6.310);
\gpcolor{color=gp lt color border}
\gpsetlinetype{gp lt border}
\gpsetdashtype{gp dt solid}
\gpsetlinewidth{1.00}
\draw[gp path] (5.211,1.366)--(5.211,1.546);
\draw[gp path] (5.211,6.310)--(5.211,6.130);
\node[gp node center] at (5.211,1.058) {$100$};
\draw[gp path] (5.583,1.366)--(5.583,1.456);
\draw[gp path] (5.583,6.310)--(5.583,6.220);
\draw[gp path] (5.801,1.366)--(5.801,1.456);
\draw[gp path] (5.801,6.310)--(5.801,6.220);
\draw[gp path] (5.955,1.366)--(5.955,1.456);
\draw[gp path] (5.955,6.310)--(5.955,6.220);
\draw[gp path] (6.075,1.366)--(6.075,1.456);
\draw[gp path] (6.075,6.310)--(6.075,6.220);
\draw[gp path] (6.173,1.366)--(6.173,1.456);
\draw[gp path] (6.173,6.310)--(6.173,6.220);
\draw[gp path] (6.256,1.366)--(6.256,1.456);
\draw[gp path] (6.256,6.310)--(6.256,6.220);
\draw[gp path] (6.327,1.366)--(6.327,1.456);
\draw[gp path] (6.327,6.310)--(6.327,6.220);
\draw[gp path] (6.390,1.366)--(6.390,1.456);
\draw[gp path] (6.390,6.310)--(6.390,6.220);
\gpcolor{color=gp lt color axes}
\gpsetlinetype{gp lt axes}
\gpsetdashtype{gp dt axes}
\gpsetlinewidth{0.50}
\draw[gp path] (6.447,1.366)--(6.447,6.310);
\gpcolor{color=gp lt color border}
\gpsetlinetype{gp lt border}
\gpsetdashtype{gp dt solid}
\gpsetlinewidth{1.00}
\draw[gp path] (6.447,1.366)--(6.447,1.546);
\draw[gp path] (6.447,6.310)--(6.447,6.130);
\node[gp node center] at (6.447,1.058) {$1000$};
\draw[gp path] (1.504,6.310)--(1.504,1.366)--(6.447,1.366)--(6.447,6.310)--cycle;
\node[gp node center,rotate=-270] at (0.292,3.838) {CPU Time with Int-Split [h]};
\node[gp node center] at (3.975,0.596) {CPU Time without Int-Split [h]};
\gpcolor{rgb color={0.000,0.000,0.000}}
\draw[gp path] (1.504,1.366)--(1.554,1.416)--(1.604,1.466)--(1.654,1.516)--(1.704,1.566)%
  --(1.754,1.616)--(1.804,1.666)--(1.854,1.716)--(1.903,1.766)--(1.953,1.815)--(2.003,1.865)%
  --(2.053,1.915)--(2.103,1.965)--(2.153,2.015)--(2.203,2.065)--(2.253,2.115)--(2.303,2.165)%
  --(2.353,2.215)--(2.403,2.265)--(2.453,2.315)--(2.503,2.365)--(2.553,2.415)--(2.602,2.465)%
  --(2.652,2.515)--(2.702,2.565)--(2.752,2.614)--(2.802,2.664)--(2.852,2.714)--(2.902,2.764)%
  --(2.952,2.814)--(3.002,2.864)--(3.052,2.914)--(3.102,2.964)--(3.152,3.014)--(3.202,3.064)%
  --(3.252,3.114)--(3.301,3.164)--(3.351,3.214)--(3.401,3.264)--(3.451,3.314)--(3.501,3.364)%
  --(3.551,3.414)--(3.601,3.463)--(3.651,3.513)--(3.701,3.563)--(3.751,3.613)--(3.801,3.663)%
  --(3.851,3.713)--(3.901,3.763)--(3.951,3.813)--(4.000,3.863)--(4.050,3.913)--(4.100,3.963)%
  --(4.150,4.013)--(4.200,4.063)--(4.250,4.113)--(4.300,4.163)--(4.350,4.213)--(4.400,4.262)%
  --(4.450,4.312)--(4.500,4.362)--(4.550,4.412)--(4.600,4.462)--(4.650,4.512)--(4.699,4.562)%
  --(4.749,4.612)--(4.799,4.662)--(4.849,4.712)--(4.899,4.762)--(4.949,4.812)--(4.999,4.862)%
  --(5.049,4.912)--(5.099,4.962)--(5.149,5.012)--(5.199,5.062)--(5.249,5.111)--(5.299,5.161)%
  --(5.349,5.211)--(5.398,5.261)--(5.448,5.311)--(5.498,5.361)--(5.548,5.411)--(5.598,5.461)%
  --(5.648,5.511)--(5.698,5.561)--(5.748,5.611)--(5.798,5.661)--(5.848,5.711)--(5.898,5.761)%
  --(5.948,5.811)--(5.998,5.861)--(6.048,5.910)--(6.097,5.960)--(6.147,6.010)--(6.197,6.060)%
  --(6.247,6.110)--(6.297,6.160)--(6.347,6.210)--(6.397,6.260)--(6.447,6.310);
\gpsetpointsize{4.00}
\gp3point{gp mark 2}{}{(1.847,1.560)}
\gp3point{gp mark 2}{}{(2.129,1.889)}
\gp3point{gp mark 2}{}{(2.568,2.338)}
\gp3point{gp mark 2}{}{(4.272,4.041)}
\gp3point{gp mark 2}{}{(5.074,4.876)}
\gp3point{gp mark 2}{}{(5.157,4.971)}
\gp3point{gp mark 2}{}{(6.243,5.981)}
\gp3point{gp mark 2}{}{(6.270,6.013)}
\gp3point{gp mark 2}{}{(6.289,6.022)}
\gp3point{gp mark 2}{}{(6.342,6.072)}
\gp3point{gp mark 2}{}{(6.342,6.072)}
\gp3point{gp mark 2}{}{(3.888,3.410)}
\gp3point{gp mark 2}{}{(4.011,3.660)}
\gp3point{gp mark 2}{}{(4.182,3.740)}
\gp3point{gp mark 2}{}{(4.347,3.860)}
\gp3point{gp mark 2}{}{(5.082,4.908)}
\gp3point{gp mark 2}{}{(6.357,6.166)}
\gp3point{gp mark 2}{}{(6.357,6.178)}
\gp3point{gp mark 2}{}{(6.445,6.289)}
\gpcolor{color=gp lt color border}
\draw[gp path] (1.504,6.310)--(1.504,1.366)--(6.447,1.366)--(6.447,6.310)--cycle;
%% coordinates of the plot area
\gpdefrectangularnode{gp plot 1}{\pgfpoint{1.504cm}{1.366cm}}{\pgfpoint{6.447cm}{6.310cm}}
\end{tikzpicture}
%% gnuplot variables

%% file: sn-gttt.tex
\begin{tikzpicture}[gnuplot]
%% generated with GNUPLOT 5.4p1 (Lua 5.3; terminal rev. Jun 2020, script rev. 114)
%% Mon Mar 20 13:36:36 2023
\gpcolor{color=gp lt color axes}
\gpsetlinetype{gp lt axes}
\gpsetdashtype{gp dt axes}
\gpsetlinewidth{0.50}
\draw[gp path] (1.320,0.616)--(14.447,0.616);
\gpcolor{color=gp lt color border}
\gpsetlinetype{gp lt border}
\gpsetdashtype{gp dt solid}
\gpsetlinewidth{1.00}
\draw[gp path] (1.320,0.616)--(1.500,0.616);
\draw[gp path] (14.447,0.616)--(14.267,0.616);
\node[gp node right] at (1.136,0.616) {$0$};
\gpcolor{color=gp lt color axes}
\gpsetlinetype{gp lt axes}
\gpsetdashtype{gp dt axes}
\gpsetlinewidth{0.50}
\draw[gp path] (1.320,1.031)--(14.447,1.031);
\gpcolor{color=gp lt color border}
\gpsetlinetype{gp lt border}
\gpsetdashtype{gp dt solid}
\gpsetlinewidth{1.00}
\draw[gp path] (1.320,1.031)--(1.500,1.031);
\draw[gp path] (14.447,1.031)--(14.267,1.031);
\node[gp node right] at (1.136,1.031) {$20$};
\gpcolor{color=gp lt color axes}
\gpsetlinetype{gp lt axes}
\gpsetdashtype{gp dt axes}
\gpsetlinewidth{0.50}
\draw[gp path] (1.320,1.446)--(14.447,1.446);
\gpcolor{color=gp lt color border}
\gpsetlinetype{gp lt border}
\gpsetdashtype{gp dt solid}
\gpsetlinewidth{1.00}
\draw[gp path] (1.320,1.446)--(1.500,1.446);
\draw[gp path] (14.447,1.446)--(14.267,1.446);
\node[gp node right] at (1.136,1.446) {$40$};
\gpcolor{color=gp lt color axes}
\gpsetlinetype{gp lt axes}
\gpsetdashtype{gp dt axes}
\gpsetlinewidth{0.50}
\draw[gp path] (1.320,1.861)--(14.447,1.861);
\gpcolor{color=gp lt color border}
\gpsetlinetype{gp lt border}
\gpsetdashtype{gp dt solid}
\gpsetlinewidth{1.00}
\draw[gp path] (1.320,1.861)--(1.500,1.861);
\draw[gp path] (14.447,1.861)--(14.267,1.861);
\node[gp node right] at (1.136,1.861) {$60$};
\gpcolor{color=gp lt color axes}
\gpsetlinetype{gp lt axes}
\gpsetdashtype{gp dt axes}
\gpsetlinewidth{0.50}
\draw[gp path] (1.320,2.276)--(1.504,2.276);
\draw[gp path] (4.996,2.276)--(14.447,2.276);
\gpcolor{color=gp lt color border}
\gpsetlinetype{gp lt border}
\gpsetdashtype{gp dt solid}
\gpsetlinewidth{1.00}
\draw[gp path] (1.320,2.276)--(1.500,2.276);
\draw[gp path] (14.447,2.276)--(14.267,2.276);
\node[gp node right] at (1.136,2.276) {$80$};
\gpcolor{color=gp lt color axes}
\gpsetlinetype{gp lt axes}
\gpsetdashtype{gp dt axes}
\gpsetlinewidth{0.50}
\draw[gp path] (1.320,2.691)--(14.447,2.691);
\gpcolor{color=gp lt color border}
\gpsetlinetype{gp lt border}
\gpsetdashtype{gp dt solid}
\gpsetlinewidth{1.00}
\draw[gp path] (1.320,2.691)--(1.500,2.691);
\draw[gp path] (14.447,2.691)--(14.267,2.691);
\node[gp node right] at (1.136,2.691) {$100$};
\gpcolor{color=gp lt color axes}
\gpsetlinetype{gp lt axes}
\gpsetdashtype{gp dt axes}
\gpsetlinewidth{0.50}
\draw[gp path] (1.917,0.616)--(1.917,1.895);
\draw[gp path] (1.917,2.511)--(1.917,2.691);
\gpcolor{color=gp lt color border}
\gpsetlinetype{gp lt border}
\gpsetdashtype{gp dt solid}
\gpsetlinewidth{1.00}
\draw[gp path] (1.917,0.616)--(1.917,0.796);
\draw[gp path] (1.917,2.691)--(1.917,2.511);
\node[gp node center] at (1.917,0.308) {domino};
\gpcolor{color=gp lt color axes}
\gpsetlinetype{gp lt axes}
\gpsetdashtype{gp dt axes}
\gpsetlinewidth{0.50}
\draw[gp path] (3.110,0.616)--(3.110,1.895);
\draw[gp path] (3.110,2.511)--(3.110,2.691);
\gpcolor{color=gp lt color border}
\gpsetlinetype{gp lt border}
\gpsetdashtype{gp dt solid}
\gpsetlinewidth{1.00}
\draw[gp path] (3.110,0.616)--(3.110,0.796);
\draw[gp path] (3.110,2.691)--(3.110,2.511);
\node[gp node center] at (3.110,0.308) {el};
\gpcolor{color=gp lt color axes}
\gpsetlinetype{gp lt axes}
\gpsetdashtype{gp dt axes}
\gpsetlinewidth{0.50}
\draw[gp path] (4.303,0.616)--(4.303,1.895);
\draw[gp path] (4.303,2.511)--(4.303,2.691);
\gpcolor{color=gp lt color border}
\gpsetlinetype{gp lt border}
\gpsetdashtype{gp dt solid}
\gpsetlinewidth{1.00}
\draw[gp path] (4.303,0.616)--(4.303,0.796);
\draw[gp path] (4.303,2.691)--(4.303,2.511);
\node[gp node center] at (4.303,0.308) {tic};
\gpcolor{color=gp lt color axes}
\gpsetlinetype{gp lt axes}
\gpsetdashtype{gp dt axes}
\gpsetlinewidth{0.50}
\draw[gp path] (5.497,0.616)--(5.497,2.691);
\gpcolor{color=gp lt color border}
\gpsetlinetype{gp lt border}
\gpsetdashtype{gp dt solid}
\gpsetlinewidth{1.00}
\draw[gp path] (5.497,0.616)--(5.497,0.796);
\draw[gp path] (5.497,2.691)--(5.497,2.511);
\node[gp node center] at (5.497,0.308) {elly};
\gpcolor{color=gp lt color axes}
\gpsetlinetype{gp lt axes}
\gpsetdashtype{gp dt axes}
\gpsetlinewidth{0.50}
\draw[gp path] (6.690,0.616)--(6.690,2.691);
\gpcolor{color=gp lt color border}
\gpsetlinetype{gp lt border}
\gpsetdashtype{gp dt solid}
\gpsetlinewidth{1.00}
\draw[gp path] (6.690,0.616)--(6.690,0.796);
\draw[gp path] (6.690,2.691)--(6.690,2.511);
\node[gp node center] at (6.690,0.308) {tippy};
\gpcolor{color=gp lt color axes}
\gpsetlinetype{gp lt axes}
\gpsetdashtype{gp dt axes}
\gpsetlinewidth{0.50}
\draw[gp path] (7.884,0.616)--(7.884,2.691);
\gpcolor{color=gp lt color border}
\gpsetlinetype{gp lt border}
\gpsetdashtype{gp dt solid}
\gpsetlinewidth{1.00}
\draw[gp path] (7.884,0.616)--(7.884,0.796);
\draw[gp path] (7.884,2.691)--(7.884,2.511);
\node[gp node center] at (7.884,0.308) {knobby};
\gpcolor{color=gp lt color axes}
\gpsetlinetype{gp lt axes}
\gpsetdashtype{gp dt axes}
\gpsetlinewidth{0.50}
\draw[gp path] (9.077,0.616)--(9.077,2.691);
\gpcolor{color=gp lt color border}
\gpsetlinetype{gp lt border}
\gpsetdashtype{gp dt solid}
\gpsetlinewidth{1.00}
\draw[gp path] (9.077,0.616)--(9.077,0.796);
\draw[gp path] (9.077,2.691)--(9.077,2.511);
\node[gp node center] at (9.077,0.308) {skinny};
\gpcolor{color=gp lt color axes}
\gpsetlinetype{gp lt axes}
\gpsetdashtype{gp dt axes}
\gpsetlinewidth{0.50}
\draw[gp path] (10.270,0.616)--(10.270,2.691);
\gpcolor{color=gp lt color border}
\gpsetlinetype{gp lt border}
\gpsetdashtype{gp dt solid}
\gpsetlinewidth{1.00}
\draw[gp path] (10.270,0.616)--(10.270,0.796);
\draw[gp path] (10.270,2.691)--(10.270,2.511);
\node[gp node center] at (10.270,0.308) {Z};
\gpcolor{color=gp lt color axes}
\gpsetlinetype{gp lt axes}
\gpsetdashtype{gp dt axes}
\gpsetlinewidth{0.50}
\draw[gp path] (11.464,0.616)--(11.464,2.691);
\gpcolor{color=gp lt color border}
\gpsetlinetype{gp lt border}
\gpsetdashtype{gp dt solid}
\gpsetlinewidth{1.00}
\draw[gp path] (11.464,0.616)--(11.464,0.796);
\draw[gp path] (11.464,2.691)--(11.464,2.511);
\node[gp node center] at (11.464,0.308) {L};
\gpcolor{color=gp lt color axes}
\gpsetlinetype{gp lt axes}
\gpsetdashtype{gp dt axes}
\gpsetlinewidth{0.50}
\draw[gp path] (12.657,0.616)--(12.657,2.691);
\gpcolor{color=gp lt color border}
\gpsetlinetype{gp lt border}
\gpsetdashtype{gp dt solid}
\gpsetlinewidth{1.00}
\draw[gp path] (12.657,0.616)--(12.657,0.796);
\draw[gp path] (12.657,2.691)--(12.657,2.511);
\node[gp node center] at (12.657,0.308) {fatty};
\gpcolor{color=gp lt color axes}
\gpsetlinetype{gp lt axes}
\gpsetdashtype{gp dt axes}
\gpsetlinewidth{0.50}
\draw[gp path] (13.850,0.616)--(13.850,2.691);
\gpcolor{color=gp lt color border}
\gpsetlinetype{gp lt border}
\gpsetdashtype{gp dt solid}
\gpsetlinewidth{1.00}
\draw[gp path] (13.850,0.616)--(13.850,0.796);
\draw[gp path] (13.850,2.691)--(13.850,2.511);
\node[gp node center] at (13.850,0.308) {snaky};
\draw[gp path] (1.320,2.691)--(1.320,0.616)--(14.447,0.616)--(14.447,2.691)--cycle;
\node[gp node center,rotate=-270] at (0.292,1.653) {CPU Time [h]};
\node[gp node right] at (3.712,2.357) {No Int-Split};
\gpfill{rgb color={0.071,0.737,0.988}} (3.896,2.280)--(4.812,2.280)--(4.812,2.434)--(3.896,2.434)--cycle;
\gpcolor{rgb color={0.071,0.737,0.988}}
\draw[gp path] (3.896,2.280)--(4.812,2.280)--(4.812,2.434)--(3.896,2.434)--cycle;
\gpfill{rgb color={0.071,0.737,0.988}} (1.519,0.616)--(1.918,0.616)--(1.918,0.621)--(1.519,0.621)--cycle;
\draw[gp path] (1.519,0.616)--(1.519,0.620)--(1.917,0.620)--(1.917,0.616)--cycle;
\gpfill{rgb color={0.071,0.737,0.988}} (2.712,0.616)--(3.111,0.616)--(3.111,0.624)--(2.712,0.624)--cycle;
\draw[gp path] (2.712,0.616)--(2.712,0.623)--(3.110,0.623)--(3.110,0.616)--cycle;
\gpfill{rgb color={0.071,0.737,0.988}} (3.906,0.616)--(4.304,0.616)--(4.304,0.632)--(3.906,0.632)--cycle;
\draw[gp path] (3.906,0.616)--(3.906,0.631)--(4.303,0.631)--(4.303,0.616)--cycle;
\gpfill{rgb color={0.071,0.737,0.988}} (5.099,0.616)--(5.498,0.616)--(5.498,0.977)--(5.099,0.977)--cycle;
\draw[gp path] (5.099,0.616)--(5.099,0.976)--(5.497,0.976)--(5.497,0.616)--cycle;
\gpfill{rgb color={0.071,0.737,0.988}} (6.292,0.616)--(6.691,0.616)--(6.691,2.224)--(6.292,2.224)--cycle;
\draw[gp path] (6.292,0.616)--(6.292,2.223)--(6.690,2.223)--(6.690,0.616)--cycle;
\gpfill{rgb color={0.071,0.737,0.988}} (7.486,0.616)--(7.885,0.616)--(7.885,2.492)--(7.486,2.492)--cycle;
\draw[gp path] (7.486,0.616)--(7.486,2.491)--(7.884,2.491)--(7.884,0.616)--cycle;
\gpfill{rgb color={0.071,0.737,0.988}} (8.679,0.616)--(9.078,0.616)--(9.078,2.692)--(8.679,2.692)--cycle;
\draw[gp path] (8.679,0.616)--(8.679,2.691)--(9.077,2.691)--(9.077,0.616)--cycle;
\gpfill{rgb color={0.071,0.737,0.988}} (9.872,0.616)--(10.271,0.616)--(10.271,2.692)--(9.872,2.692)--cycle;
\draw[gp path] (9.872,0.616)--(9.872,2.691)--(10.270,2.691)--(10.270,0.616)--cycle;
\gpfill{rgb color={0.071,0.737,0.988}} (11.066,0.616)--(11.465,0.616)--(11.465,2.692)--(11.066,2.692)--cycle;
\draw[gp path] (11.066,0.616)--(11.066,2.691)--(11.464,2.691)--(11.464,0.616)--cycle;
\gpfill{rgb color={0.071,0.737,0.988}} (12.259,0.616)--(12.658,0.616)--(12.658,2.692)--(12.259,2.692)--cycle;
\draw[gp path] (12.259,0.616)--(12.259,2.691)--(12.657,2.691)--(12.657,0.616)--cycle;
\gpfill{rgb color={0.071,0.737,0.988}} (13.453,0.616)--(13.851,0.616)--(13.851,2.692)--(13.453,2.692)--cycle;
\draw[gp path] (13.453,0.616)--(13.453,2.691)--(13.850,2.691)--(13.850,0.616)--cycle;
\gpcolor{color=gp lt color border}
\node[gp node center,font={\fontsize{1.5pt}{1.8pt}\selectfont}] at (1.718,0.744) {0.69};
\node[gp node center,font={\fontsize{1.5pt}{1.8pt}\selectfont}] at (2.911,0.747) {0.82};
\node[gp node center,font={\fontsize{1.5pt}{1.8pt}\selectfont}] at (4.105,0.756) {1.23};
\node[gp node center,font={\fontsize{1.5pt}{1.8pt}\selectfont}] at (5.298,1.101) { 18};
\node[gp node center,font={\fontsize{1.5pt}{1.8pt}\selectfont}] at (6.491,2.347) { 78};
\node[gp node center,font={\fontsize{1.5pt}{1.8pt}\selectfont}] at (7.685,2.442) { 91};
\node[gp node center,font={\fontsize{1.5pt}{1.8pt}\selectfont}] at (8.878,2.442) {684};
\node[gp node center,font={\fontsize{1.5pt}{1.8pt}\selectfont}] at (10.071,2.442) {719};
\node[gp node center,font={\fontsize{1.5pt}{1.8pt}\selectfont}] at (11.265,2.442) {745};
\node[gp node center,font={\fontsize{1.5pt}{1.8pt}\selectfont}] at (12.458,2.442) {822};
\node[gp node center,font={\fontsize{1.5pt}{1.8pt}\selectfont}] at (13.651,2.442) {823};
\node[gp node right] at (3.712,2.049) {Int-Split};
\gpfill{rgb color={0.988,0.780,0.071}} (3.896,1.972)--(4.812,1.972)--(4.812,2.126)--(3.896,2.126)--cycle;
\gpcolor{rgb color={0.988,0.780,0.071}}
\draw[gp path] (3.896,1.972)--(4.812,1.972)--(4.812,2.126)--(3.896,2.126)--cycle;
\gpfill{rgb color={0.988,0.780,0.071}} (1.917,0.616)--(2.315,0.616)--(2.315,0.620)--(1.917,0.620)--cycle;
\draw[gp path] (1.917,0.616)--(1.917,0.619)--(2.314,0.619)--(2.314,0.616)--cycle;
\gpfill{rgb color={0.988,0.780,0.071}} (3.110,0.616)--(3.509,0.616)--(3.509,0.623)--(3.110,0.623)--cycle;
\draw[gp path] (3.110,0.616)--(3.110,0.622)--(3.508,0.622)--(3.508,0.616)--cycle;
\gpfill{rgb color={0.988,0.780,0.071}} (4.303,0.616)--(4.702,0.616)--(4.702,0.630)--(4.303,0.630)--cycle;
\draw[gp path] (4.303,0.616)--(4.303,0.629)--(4.701,0.629)--(4.701,0.616)--cycle;
\gpfill{rgb color={0.988,0.780,0.071}} (5.497,0.616)--(5.896,0.616)--(5.896,0.920)--(5.497,0.920)--cycle;
\draw[gp path] (5.497,0.616)--(5.497,0.919)--(5.895,0.919)--(5.895,0.616)--cycle;
\gpfill{rgb color={0.988,0.780,0.071}} (6.690,0.616)--(7.089,0.616)--(7.089,2.053)--(6.690,2.053)--cycle;
\draw[gp path] (6.690,0.616)--(6.690,2.052)--(7.088,2.052)--(7.088,0.616)--cycle;
\gpfill{rgb color={0.988,0.780,0.071}} (7.884,0.616)--(8.282,0.616)--(8.282,2.331)--(7.884,2.331)--cycle;
\draw[gp path] (7.884,0.616)--(7.884,2.330)--(8.281,2.330)--(8.281,0.616)--cycle;
\gpfill{rgb color={0.988,0.780,0.071}} (9.077,0.616)--(9.476,0.616)--(9.476,2.692)--(9.077,2.692)--cycle;
\draw[gp path] (9.077,0.616)--(9.077,2.691)--(9.475,2.691)--(9.475,0.616)--cycle;
\gpfill{rgb color={0.988,0.780,0.071}} (10.270,0.616)--(10.669,0.616)--(10.669,2.692)--(10.270,2.692)--cycle;
\draw[gp path] (10.270,0.616)--(10.270,2.691)--(10.668,2.691)--(10.668,0.616)--cycle;
\gpfill{rgb color={0.988,0.780,0.071}} (11.464,0.616)--(11.862,0.616)--(11.862,2.692)--(11.464,2.692)--cycle;
\draw[gp path] (11.464,0.616)--(11.464,2.691)--(11.861,2.691)--(11.861,0.616)--cycle;
\gpfill{rgb color={0.988,0.780,0.071}} (12.657,0.616)--(13.056,0.616)--(13.056,2.692)--(12.657,2.692)--cycle;
\draw[gp path] (12.657,0.616)--(12.657,2.691)--(13.055,2.691)--(13.055,0.616)--cycle;
\gpfill{rgb color={0.988,0.780,0.071}} (13.850,0.616)--(14.249,0.616)--(14.249,2.692)--(13.850,2.692)--cycle;
\draw[gp path] (13.850,0.616)--(13.850,2.691)--(14.248,2.691)--(14.248,0.616)--cycle;
\gpcolor{color=gp lt color border}
\node[gp node center,font={\fontsize{1.5pt}{1.8pt}\selectfont}] at (2.116,0.743) {0.64};
\node[gp node center,font={\fontsize{1.5pt}{1.8pt}\selectfont}] at (3.309,0.746) {0.77};
\node[gp node center,font={\fontsize{1.5pt}{1.8pt}\selectfont}] at (4.502,0.753) {1.11};
\node[gp node center,font={\fontsize{1.5pt}{1.8pt}\selectfont}] at (5.696,1.043) { 15};
\node[gp node center,font={\fontsize{1.5pt}{1.8pt}\selectfont}] at (6.889,2.176) { 70};
\node[gp node center,font={\fontsize{1.5pt}{1.8pt}\selectfont}] at (8.082,2.442) { 83};
\node[gp node center,font={\fontsize{1.5pt}{1.8pt}\selectfont}] at (9.276,2.442) {542};
\node[gp node center,font={\fontsize{1.5pt}{1.8pt}\selectfont}] at (10.469,2.442) {575};
\node[gp node center,font={\fontsize{1.5pt}{1.8pt}\selectfont}] at (11.662,2.442) {586};
\node[gp node center,font={\fontsize{1.5pt}{1.8pt}\selectfont}] at (12.856,2.442) {643};
\node[gp node center,font={\fontsize{1.5pt}{1.8pt}\selectfont}] at (14.049,2.442) {643};
\draw[gp path] (1.320,2.691)--(1.320,0.616)--(14.447,0.616)--(14.447,2.691)--cycle;
%% coordinates of the plot area
\gpdefrectangularnode{gp plot 1}{\pgfpoint{1.320cm}{0.616cm}}{\pgfpoint{14.447cm}{2.691cm}}
\end{tikzpicture}
%% gnuplot variables

%% file: sn-hex.tex
\begin{tikzpicture}[gnuplot]
%% generated with GNUPLOT 5.4p1 (Lua 5.3; terminal rev. Jun 2020, script rev. 114)
%% Mon Mar 20 13:37:17 2023
\gpcolor{color=gp lt color axes}
\gpsetlinetype{gp lt axes}
\gpsetdashtype{gp dt axes}
\gpsetlinewidth{0.50}
\draw[gp path] (1.320,0.616)--(14.447,0.616);
\gpcolor{color=gp lt color border}
\gpsetlinetype{gp lt border}
\gpsetdashtype{gp dt solid}
\gpsetlinewidth{1.00}
\draw[gp path] (1.320,0.616)--(1.500,0.616);
\draw[gp path] (14.447,0.616)--(14.267,0.616);
\node[gp node right] at (1.136,0.616) {$0$};
\gpcolor{color=gp lt color axes}
\gpsetlinetype{gp lt axes}
\gpsetdashtype{gp dt axes}
\gpsetlinewidth{0.50}
\draw[gp path] (1.320,1.031)--(14.447,1.031);
\gpcolor{color=gp lt color border}
\gpsetlinetype{gp lt border}
\gpsetdashtype{gp dt solid}
\gpsetlinewidth{1.00}
\draw[gp path] (1.320,1.031)--(1.500,1.031);
\draw[gp path] (14.447,1.031)--(14.267,1.031);
\node[gp node right] at (1.136,1.031) {$20$};
\gpcolor{color=gp lt color axes}
\gpsetlinetype{gp lt axes}
\gpsetdashtype{gp dt axes}
\gpsetlinewidth{0.50}
\draw[gp path] (1.320,1.446)--(14.447,1.446);
\gpcolor{color=gp lt color border}
\gpsetlinetype{gp lt border}
\gpsetdashtype{gp dt solid}
\gpsetlinewidth{1.00}
\draw[gp path] (1.320,1.446)--(1.500,1.446);
\draw[gp path] (14.447,1.446)--(14.267,1.446);
\node[gp node right] at (1.136,1.446) {$40$};
\gpcolor{color=gp lt color axes}
\gpsetlinetype{gp lt axes}
\gpsetdashtype{gp dt axes}
\gpsetlinewidth{0.50}
\draw[gp path] (1.320,1.861)--(14.447,1.861);
\gpcolor{color=gp lt color border}
\gpsetlinetype{gp lt border}
\gpsetdashtype{gp dt solid}
\gpsetlinewidth{1.00}
\draw[gp path] (1.320,1.861)--(1.500,1.861);
\draw[gp path] (14.447,1.861)--(14.267,1.861);
\node[gp node right] at (1.136,1.861) {$60$};
\gpcolor{color=gp lt color axes}
\gpsetlinetype{gp lt axes}
\gpsetdashtype{gp dt axes}
\gpsetlinewidth{0.50}
\draw[gp path] (1.320,2.276)--(14.447,2.276);
\gpcolor{color=gp lt color border}
\gpsetlinetype{gp lt border}
\gpsetdashtype{gp dt solid}
\gpsetlinewidth{1.00}
\draw[gp path] (1.320,2.276)--(1.500,2.276);
\draw[gp path] (14.447,2.276)--(14.267,2.276);
\node[gp node right] at (1.136,2.276) {$80$};
\gpcolor{color=gp lt color axes}
\gpsetlinetype{gp lt axes}
\gpsetdashtype{gp dt axes}
\gpsetlinewidth{0.50}
\draw[gp path] (1.320,2.691)--(14.447,2.691);
\gpcolor{color=gp lt color border}
\gpsetlinetype{gp lt border}
\gpsetdashtype{gp dt solid}
\gpsetlinewidth{1.00}
\draw[gp path] (1.320,2.691)--(1.500,2.691);
\draw[gp path] (14.447,2.691)--(14.267,2.691);
\node[gp node right] at (1.136,2.691) {$100$};
\gpcolor{color=gp lt color axes}
\gpsetlinetype{gp lt axes}
\gpsetdashtype{gp dt axes}
\gpsetlinewidth{0.50}
\draw[gp path] (2.140,0.616)--(2.140,2.691);
\gpcolor{color=gp lt color border}
\gpsetlinetype{gp lt border}
\gpsetdashtype{gp dt solid}
\gpsetlinewidth{1.00}
\draw[gp path] (2.140,0.616)--(2.140,0.796);
\draw[gp path] (2.140,2.691)--(2.140,2.511);
\node[gp node center] at (2.140,0.308) {10 5x5};
\gpcolor{color=gp lt color axes}
\gpsetlinetype{gp lt axes}
\gpsetdashtype{gp dt axes}
\gpsetlinewidth{0.50}
\draw[gp path] (3.781,0.616)--(3.781,2.691);
\gpcolor{color=gp lt color border}
\gpsetlinetype{gp lt border}
\gpsetdashtype{gp dt solid}
\gpsetlinewidth{1.00}
\draw[gp path] (3.781,0.616)--(3.781,0.796);
\draw[gp path] (3.781,2.691)--(3.781,2.511);
\node[gp node center] at (3.781,0.308) {16 5x5};
\gpcolor{color=gp lt color axes}
\gpsetlinetype{gp lt axes}
\gpsetdashtype{gp dt axes}
\gpsetlinewidth{0.50}
\draw[gp path] (5.422,0.616)--(5.422,2.691);
\gpcolor{color=gp lt color border}
\gpsetlinetype{gp lt border}
\gpsetdashtype{gp dt solid}
\gpsetlinewidth{1.00}
\draw[gp path] (5.422,0.616)--(5.422,0.796);
\draw[gp path] (5.422,2.691)--(5.422,2.511);
\node[gp node center] at (5.422,0.308) {02 5x5};
\gpcolor{color=gp lt color axes}
\gpsetlinetype{gp lt axes}
\gpsetdashtype{gp dt axes}
\gpsetlinewidth{0.50}
\draw[gp path] (7.063,0.616)--(7.063,2.691);
\gpcolor{color=gp lt color border}
\gpsetlinetype{gp lt border}
\gpsetdashtype{gp dt solid}
\gpsetlinewidth{1.00}
\draw[gp path] (7.063,0.616)--(7.063,0.796);
\draw[gp path] (7.063,2.691)--(7.063,2.511);
\node[gp node center] at (7.063,0.308) {15 5x5};
\gpcolor{color=gp lt color axes}
\gpsetlinetype{gp lt axes}
\gpsetdashtype{gp dt axes}
\gpsetlinewidth{0.50}
\draw[gp path] (8.704,0.616)--(8.704,2.691);
\gpcolor{color=gp lt color border}
\gpsetlinetype{gp lt border}
\gpsetdashtype{gp dt solid}
\gpsetlinewidth{1.00}
\draw[gp path] (8.704,0.616)--(8.704,0.796);
\draw[gp path] (8.704,2.691)--(8.704,2.511);
\node[gp node center] at (8.704,0.308) {05 6x6};
\gpcolor{color=gp lt color axes}
\gpsetlinetype{gp lt axes}
\gpsetdashtype{gp dt axes}
\gpsetlinewidth{0.50}
\draw[gp path] (10.345,0.616)--(10.345,2.691);
\gpcolor{color=gp lt color border}
\gpsetlinetype{gp lt border}
\gpsetdashtype{gp dt solid}
\gpsetlinewidth{1.00}
\draw[gp path] (10.345,0.616)--(10.345,0.796);
\draw[gp path] (10.345,2.691)--(10.345,2.511);
\node[gp node center] at (10.345,0.308) {03 6x6};
\gpcolor{color=gp lt color axes}
\gpsetlinetype{gp lt axes}
\gpsetdashtype{gp dt axes}
\gpsetlinewidth{0.50}
\draw[gp path] (11.986,0.616)--(11.986,2.691);
\gpcolor{color=gp lt color border}
\gpsetlinetype{gp lt border}
\gpsetdashtype{gp dt solid}
\gpsetlinewidth{1.00}
\draw[gp path] (11.986,0.616)--(11.986,0.796);
\draw[gp path] (11.986,2.691)--(11.986,2.511);
\node[gp node center] at (11.986,0.308) {17 6x6};
\gpcolor{color=gp lt color axes}
\gpsetlinetype{gp lt axes}
\gpsetdashtype{gp dt axes}
\gpsetlinewidth{0.50}
\draw[gp path] (13.627,0.616)--(13.627,2.691);
\gpcolor{color=gp lt color border}
\gpsetlinetype{gp lt border}
\gpsetdashtype{gp dt solid}
\gpsetlinewidth{1.00}
\draw[gp path] (13.627,0.616)--(13.627,0.796);
\draw[gp path] (13.627,2.691)--(13.627,2.511);
\node[gp node center] at (13.627,0.308) {20 6x6};
\draw[gp path] (1.320,2.691)--(1.320,0.616)--(14.447,0.616)--(14.447,2.691)--cycle;
\node[gp node center,rotate=-270] at (0.292,1.653) {CPU Time [h]};
\gpfill{rgb color={0.071,0.737,0.988}} (1.593,0.616)--(2.141,0.616)--(2.141,0.793)--(1.593,0.793)--cycle;
\gpcolor{rgb color={0.071,0.737,0.988}}
\draw[gp path] (1.593,0.616)--(1.593,0.792)--(2.140,0.792)--(2.140,0.616)--cycle;
\gpfill{rgb color={0.071,0.737,0.988}} (3.234,0.616)--(3.782,0.616)--(3.782,0.839)--(3.234,0.839)--cycle;
\draw[gp path] (3.234,0.616)--(3.234,0.838)--(3.781,0.838)--(3.781,0.616)--cycle;
\gpfill{rgb color={0.071,0.737,0.988}} (4.875,0.616)--(5.423,0.616)--(5.423,0.922)--(4.875,0.922)--cycle;
\draw[gp path] (4.875,0.616)--(4.875,0.921)--(5.422,0.921)--(5.422,0.616)--cycle;
\gpfill{rgb color={0.071,0.737,0.988}} (6.516,0.616)--(7.064,0.616)--(7.064,1.032)--(6.516,1.032)--cycle;
\draw[gp path] (6.516,0.616)--(6.516,1.031)--(7.063,1.031)--(7.063,0.616)--cycle;
\gpfill{rgb color={0.071,0.737,0.988}} (8.157,0.616)--(8.705,0.616)--(8.705,2.249)--(8.157,2.249)--cycle;
\draw[gp path] (8.157,0.616)--(8.157,2.248)--(8.704,2.248)--(8.704,0.616)--cycle;
\gpfill{rgb color={0.071,0.737,0.988}} (9.798,0.616)--(10.346,0.616)--(10.346,2.692)--(9.798,2.692)--cycle;
\draw[gp path] (9.798,0.616)--(9.798,2.691)--(10.345,2.691)--(10.345,0.616)--cycle;
\gpfill{rgb color={0.071,0.737,0.988}} (11.439,0.616)--(11.987,0.616)--(11.987,2.692)--(11.439,2.692)--cycle;
\draw[gp path] (11.439,0.616)--(11.439,2.691)--(11.986,2.691)--(11.986,0.616)--cycle;
\gpfill{rgb color={0.071,0.737,0.988}} (13.080,0.616)--(13.628,0.616)--(13.628,2.692)--(13.080,2.692)--cycle;
\draw[gp path] (13.080,0.616)--(13.080,2.691)--(13.627,2.691)--(13.627,0.616)--cycle;
\gpcolor{color=gp lt color border}
\node[gp node center,font={\fontsize{1.5pt}{1.8pt}\selectfont}] at (1.867,0.917) {  9};
\node[gp node center,font={\fontsize{1.5pt}{1.8pt}\selectfont}] at (3.508,0.962) { 11};
\node[gp node center,font={\fontsize{1.5pt}{1.8pt}\selectfont}] at (5.149,1.045) { 15};
\node[gp node center,font={\fontsize{1.5pt}{1.8pt}\selectfont}] at (6.790,1.155) { 20};
\node[gp node center,font={\fontsize{1.5pt}{1.8pt}\selectfont}] at (8.430,2.372) { 79};
\node[gp node center,font={\fontsize{1.5pt}{1.8pt}\selectfont}] at (10.071,2.442) {847};
\node[gp node center,font={\fontsize{1.5pt}{1.8pt}\selectfont}] at (11.712,2.442) {846};
\node[gp node center,font={\fontsize{1.5pt}{1.8pt}\selectfont}] at (13.353,2.442) {996};
\gpfill{rgb color={0.988,0.780,0.071}} (2.140,0.616)--(2.688,0.616)--(2.688,0.711)--(2.140,0.711)--cycle;
\gpcolor{rgb color={0.988,0.780,0.071}}
\draw[gp path] (2.140,0.616)--(2.140,0.710)--(2.687,0.710)--(2.687,0.616)--cycle;
\gpfill{rgb color={0.988,0.780,0.071}} (3.781,0.616)--(4.329,0.616)--(4.329,0.766)--(3.781,0.766)--cycle;
\draw[gp path] (3.781,0.616)--(3.781,0.765)--(4.328,0.765)--(4.328,0.616)--cycle;
\gpfill{rgb color={0.988,0.780,0.071}} (5.422,0.616)--(5.970,0.616)--(5.970,0.790)--(5.422,0.790)--cycle;
\draw[gp path] (5.422,0.616)--(5.422,0.789)--(5.969,0.789)--(5.969,0.616)--cycle;
\gpfill{rgb color={0.988,0.780,0.071}} (7.063,0.616)--(7.611,0.616)--(7.611,0.833)--(7.063,0.833)--cycle;
\draw[gp path] (7.063,0.616)--(7.063,0.832)--(7.610,0.832)--(7.610,0.616)--cycle;
\gpfill{rgb color={0.988,0.780,0.071}} (8.704,0.616)--(9.252,0.616)--(9.252,2.140)--(8.704,2.140)--cycle;
\draw[gp path] (8.704,0.616)--(8.704,2.139)--(9.251,2.139)--(9.251,0.616)--cycle;
\gpfill{rgb color={0.988,0.780,0.071}} (10.345,0.616)--(10.893,0.616)--(10.893,2.692)--(10.345,2.692)--cycle;
\draw[gp path] (10.345,0.616)--(10.345,2.691)--(10.892,2.691)--(10.892,0.616)--cycle;
\gpfill{rgb color={0.988,0.780,0.071}} (11.986,0.616)--(12.534,0.616)--(12.534,2.692)--(11.986,2.692)--cycle;
\draw[gp path] (11.986,0.616)--(11.986,2.691)--(12.533,2.691)--(12.533,0.616)--cycle;
\gpfill{rgb color={0.988,0.780,0.071}} (13.627,0.616)--(14.175,0.616)--(14.175,2.692)--(13.627,2.692)--cycle;
\draw[gp path] (13.627,0.616)--(13.627,2.691)--(14.174,2.691)--(14.174,0.616)--cycle;
\gpcolor{color=gp lt color border}
\node[gp node center,font={\fontsize{1.5pt}{1.8pt}\selectfont}] at (2.414,0.834) {  5};
\node[gp node center,font={\fontsize{1.5pt}{1.8pt}\selectfont}] at (4.055,0.889) {  8};
\node[gp node center,font={\fontsize{1.5pt}{1.8pt}\selectfont}] at (5.696,0.913) {  9};
\node[gp node center,font={\fontsize{1.5pt}{1.8pt}\selectfont}] at (7.337,0.957) { 11};
\node[gp node center,font={\fontsize{1.5pt}{1.8pt}\selectfont}] at (8.977,2.263) { 74};
\node[gp node center,font={\fontsize{1.5pt}{1.8pt}\selectfont}] at (10.618,2.442) {766};
\node[gp node center,font={\fontsize{1.5pt}{1.8pt}\selectfont}] at (12.259,2.442) {783};
\node[gp node center,font={\fontsize{1.5pt}{1.8pt}\selectfont}] at (13.900,2.442) {961};
\draw[gp path] (1.320,2.691)--(1.320,0.616)--(14.447,0.616)--(14.447,2.691)--cycle;
%% coordinates of the plot area
\gpdefrectangularnode{gp plot 1}{\pgfpoint{1.320cm}{0.616cm}}{\pgfpoint{14.447cm}{2.691cm}}
\end{tikzpicture}
%% gnuplot variables